\documentclass[conference]{IEEEtran}
\usepackage{cite}
\usepackage{amsmath,amssymb,amsfonts}
\usepackage{algpseudocode}
\usepackage{graphicx}
\usepackage{textcomp}
\usepackage{braket}
\usepackage{physics}
\usepackage{xcolor,color}

\usepackage{array,ragged2e}
\newcolumntype{P}[1]{>{\RaggedRight\arraybackslash}p{#1}}

\usepackage{hyperref}

\def\BibTeX{{\rm B\kern-.05em{\sc i\kern-.025em b}\kern-.08em
    T\kern-.1667em\lower.7ex\hbox{E}\kern-.125emX}}

\makeatletter
\newcommand{\linebreakand}{%
  \end{@IEEEauthorhalign}
  \hfill\mbox{}\par
  \mbox{}\hfill\begin{@IEEEauthorhalign}
}

\title{Simulating Noisy Quantum Circuits for Cryptographic Algorithms}

\author{\IEEEauthorblockN{Sahay Harshvardhan}
\IEEEauthorblockA{\textit{Department of Computer Science} \\
\textit{Virginia Tech}\\
Blacksburg, USA \\
bhvima@vt.edu}
\and
\IEEEauthorblockN{Sanil Jain}
\IEEEauthorblockA{\textit{College of Science} \\
\textit{Virginia Tech}\\
Blacksburg, USA \\
sanilj@vt.edu}
\and
\IEEEauthorblockN{James E. McClure}
\IEEEauthorblockA{\textit{National Security Institute} \\
\textit{Virginia Tech}\\
Blacksburg, USA \\
0000-0002-5967-5120} 
  \linebreakand 
\IEEEauthorblockN{Caleb McIrvin}
\IEEEauthorblockA{\textit{Department of Computer Science} \\
\textit{Virginia Tech}\\
Blacksburg, USA \\
0009-0009-6718-459X}
\and
\IEEEauthorblockN{Ngoc Quy Tran}
\IEEEauthorblockA{\textit{Department of Computer Science} \\
\textit{Virginia Tech}\\
Blacksburg, USA \\
ngocquy@vt.edu}
}

\begin{document}

\maketitle

\begin{abstract}
The emergence of noisy intermediate-scale quantum (NISQ) computers has important consequences for cryptographic algorithms. It is theoretically well-established that key algorithms used in cybersecurity are vulnerable to quantum computers due to the fact that theoretical security guarantees, designed based on algorithmic complexity for classical computers, are not sufficient for quantum circuits. Many different quantum algorithms have been developed, which have potentially broad applications on future computing systems. However, this potential depends on the continued maturation of quantum hardware, which remains an area of active research and development. Theoretical limits provide an upper bound on the performance for algorithms. In practice, threats to encryption can only be accurately be assessed in the context of the rapidly evolving hardware and software landscape. Software co-design refers to the concurrent design of software and hardware as a way to understand the limitations of current capabilities and develop effective strategies to advance the state of the art. Since the capabilities for classical computation currently exceed quantum capabilities, quantum emulation techniques can play an important role in the co-design process. In this paper, we describe how the {\em cuQuantum} environment can support quantum algorithm co-design activities using widely-available commodity hardware. We describe how emulation techniques can be used to assess the impact of noise on algorithms of interest, and identify limitations associated with current hardware. We present our analysis in the context of areas of priority for cybersecurity and cryptography in particular since these algorithms are extraordinarily consequential for securing information in the digital world.
\end{abstract}


\section{Introduction}
Modern cryptography relies on encryption algorithms that utilize cryptographic keys generated from complex mathematical operations. The Rivest, Shamir, Adleman (RSA) and Elliptic Curve Cryptography (ECC) algorithms are widely used and are essential for secure web browsing, secure email, and secure file transfers \cite{RSA_1978,ECC_1987}. However, the introduction of quantum computers has posed a risk to the integrity of current encryption algorithms, as they are able to process information much faster than traditional computers due to their ability to perform multiple calculations in a fundamentally different way. This has been demonstrated by Peter Shor's algorithm, which can be used to break factorization-based encryption with a sufficiently large quantum computer \cite{Shor1994}. In response to this risk, several alternatives have been proposed, such as the NTRU Cryptosystem, Goldreich–Goldwasser–Halevi Cryptosystem, Ajtai–Dwork Cryptosystem, and McEliece Cryptosystem \cite{McEliece_1978,Ajtai_1996,GGH_1997,Perlner_2009}. These algorithms are designed to be resistant to quantum computing, and are expected to become increasingly important as quantum computing technology advances.

Shor's algorithm poses a threat to many cryptographic practices that are dependent on factorization and discrete logarithm problems, such as RSA, Diffie-Hellman (DH), Elliptic Curve Cryptography (ECC), and the Digital Signature Algorithm (DSA). Table \ref{tab:shor} shows that RSA encryption is unbreakable for a standard classical computer beyond 2048-bit keys, yet its security is greatly diminished when introduced to quantum computers.  Apart from Shor’s algorithm, Grover’s algorithm poses another threat to these factorization encryptions in the post-quantum era. In addition to Shor's algorithm, Grover's algorithm is capable of brute-forcing through all possible inputs to break through the key pairs of RSA, DH, ECC, and Advanced Encryption Standard (AES) encryptions at a quadratically faster speed in quantum computers than a classical computer \cite{Vaishnavi_2021}. These vulnerabilities of quantum computers for modern cryptographies necessitate the testing of more encryptions as well as the introduction of new ones that are stronger in maintaining security integrity in the post-quantum era.

\begin{table}[htbp]
\caption{Classical Computer vs. Quantum Computer -- Time Estimation to Decrypt different size RSA Encryptions}
\begin{center}
\begin{tabular}{|c|c|c|}
\hline
\cline{2-3} 
\textbf{RSA Key Size} & \textbf{\textit{Classical Computer}}& \textbf{\textit{Quantum Computer}} \\
\hline
2048& 6 x 10$^{15}$ years & 3.00 hours \\
\hline
3072& 11 x 10$^{24}$ years & 3.53 hours \\
\hline
4096& 7 x 10$^{33}$ years & 3.94 hours \\
\hline
\multicolumn{3}{l}{$^{\mathrm{*}}$Time Estimations}
\end{tabular}
\label{tab:shor}
\end{center}
\end{table}

The race to find quantum-resistant cryptography has led to the investigation of various cryptographic techniques, such as lattice-based cryptography, code-based cryptography, hash-based signatures, multivariate-based cryptography, one-time pad (OTP) cryptography, and quantum cryptography The blend of modern and new cryptographies allows for a transition to occur between classical computers and quantum computers without the immediate fear of collapsing security infrastructures. The introduction of quantum cryptography opens the door to creating encryption keys depending on the number of photons being sent to a recipient and the orientation at which the photons are positioned \cite{Prajapati_2018}. Although quantum cryptography is still within the experimental stage, the potential it holds in introducing new encryption methods represents a movement towards increasingly secure cryptographic that can coexist with the development of quantum computers.

Classical computers have experienced exponential growth in performance for the past few decades, which has been due to the success of manufacturing processes capable of making transistors incrementally smaller and smaller in size. Smaller transistors allow for the increase in transistor density within a microchip to achieve faster computers. According to Moore’s law, the density of components in an integrated circuit, including transistors, should be doubling every 18 months \cite{Schaller_1997}. Cryptographic algorithms have proven to be robust even with the considerable increases in classical computer capabilities that have resulted from this trajectory.
Although Moore’s law projections were fairly accurate since its revision in 1975, the idea of continuing to make transistors any smaller than they currently are results in the inability to cleanly cut off and on the current leading to energy dissipation and overheating of the device \cite{Lent_2000}. Quantum computers, on the other hand, rely on Hamiltonian transformations to manipulate quantum wave functions for moving qubit particles from one state to another, and thus do not rely on transistors. If scalable manufacturing processes can be developed to support quantum computers, their potential may eventually be realized for many important applications. 
Quantum computers are most typically evaluated in terms of quantum bits (qubits) that they support, which are essentially the quantum analog to a transistor in Moore's law. Theoretical assessments of algorithms are frequently considered within this context. Due to circuit noise, indications already support the idea that qubit counts are insufficient to gauge hardware capabilities. IBM has advocated for the introduction of ``quantum volume" as a way to account for performance losses due to error correction and device noise. These factors will determine the practical impact of quantum computing and the timeline over which the impact for quantum algorithms is felt by society.




The development of fully capable quantum computers will have significant consequences for society, well beyond the dilemmas that it will create for cryptography. 
However, current quantum hardware remains relatively immature and are unable to surpass the capabilities of classical computers for typical problems of interest. 
Within this context, software tools that can emulate quantum circuits are an essential tool to develop more mature quantum algorithms and understand how the maturation of quantum hardware will influence applications for these algorithms. In this paper we will evaluate the NVIDIA {\em cuQuantum} framework as a way to support quantum programming experiments. Emulation frameworks such as {\em cuQuantum} fill an essential gap by providing a way to develop quantum codes that run on  commodity hardware that is less expensive, more widely available and more reliable as compared to  
quantum devices. We describe current capabilities for {\em cuQuantum} and describe how this approach can help us to understand how the noise characteristics for future quantum hardware may influence different algorithms as quantum technologies continue to advance. 

\section{Background}

Quantum algorithms are algorithms that are designed to be run on a quantum computer, which is a model of a computer that leverages quantum mechanics. These algorithms are designed to take advantage of quantum superposition and entanglement, which are features of quantum mechanics that allow for faster calculations than those possible on classical computers, which may consider algorithm complexities that are exponential, polynomial, or even super-polynomial. The speed-up of these algorithms is the primary benefit of using a quantum computer. Theoretical work has established that quantum algorithms may be able to provide a significant advantage in comparison to classical algorithms.

Programs developed for classical computers are based on the principles of binary logic, which utilize binary digits (bits) that are capable of representing only two possible values (0 or 1). In contrast, quantum computers employ quantum logic, based on the principles of quantum mechanics, where qubits have the ability to assume a superposition of both 0 and 1 simultaneously. This allows for manipulation of data that is stored on the qubits using quantum phenomena such as quantum entanglement, superposition, and interference. As a result, classical computer code is not directly executable on quantum computing platforms and requires quantum-specific programming techniques to be implemented.

Popular open-source frameworks for quantum computing include Qiskit by IBM, Q\# and the Quantum Development Kit by Microsoft, and Cirq by Google. These frameworks provide developers with the tools necessary to create quantum computing applications that can be deployed on various hardware platforms. They offer a range of features such as quantum circuit compilation, optimization, and simulation, as well as access to quantum hardware and cloud services. 

Quantum advantage is achieved when a quantum algorithm outperforms its classical counterpart in terms of computational speed. This is evaluated by comparing the performance of the quantum algorithm to the performance of the classical algorithm on the same problem, as the size of the input increases. To do this, the time it takes for each algorithm to complete the task is measured on simulated quantum circuits. Due to the current limitations of quantum hardware, simulations are preferred over actual quantum hardware when testing large algorithms. The comparison of the performance of the quantum and classical algorithms allows for the determination of the quantum advantage.


\begin{table*}[t]
\caption{Summary of quantum algorithm types}
\begin{center}
\begin{tabular}{|P{2.5cm}|P{5cm}|P{3cm}|P{3.6cm}|}
\hline
\textbf{Algorithm}&\multicolumn{3}{|c|}{\textbf{Characteristics}} \\
\cline{2-4} 
\textbf{Name} & \textbf{\textit{Summary}}& \textbf{\textit{Example Algorithm}}& \textbf{\textit{Time Complexity}} \\
\hline
{Algebraic and Number Theoretic} & Involves the finding of a specific number through a specialized mathematical process. & Shor's Algorithm & Polynomial \\
\hline
Oracular & Involves searching for a specific object or a set of objects. & Grover’s Algorithm & Polynomial / Superpolynomial \\
\hline
Optimization and Numerics & The calculation of a specific state given a set of constraints & Differential Equation Solver & Polynomial/ Superpolynomial  \\
\hline
Machine Learning & The calculation of a curve based on a set of observations. & Principal Component Analysis & Varies \\
\hline
\end{tabular}
\label{tab:algorithm-summary}
\end{center}
\end{table*}

Table \ref{tab:algorithm-summary} summarizes five major classes of quantum algorithms that can used to solve a plethora of problems. Algebraic and Number Theoretic Algorithms specifically are used to speed up a mathematical operation such as a Fourier Transform. These algorithms have applications in all systems. Oracular Algorithms on the other hand are involved in searching for a specific object or set of objects based on some constraints. These algorithms are run in super-polynomial time but are still relatively faster than other kinds of algorithms due to the quantum nature of the system. Approximation and Simulation Algorithms are the largest in terms of the space that is being calculated. This is due to the algorithms usually involving a random walk portion where they need to calculate all states at a point in space. This calculation is done to find all possibilities from which a point can move from point A to point B. Lastly, Optimization and Numeric Algorithms are used when solving optimization problems like Dynamic Programming. In contrast, Machine Learning is a bit of an exception to these rules as these algorithms don’t fit into any other category. A special use case that is quite interesting is quantum sorting, a kind of sorting that uses the applications of quantum mechanics and oracular algorithms. These algorithms have deep applications in cryptography and systems \cite{Jordan_2022}.  \\

\subsection{Summary of NISQ devices and challenges} 

While it has been shown theoretically that fault-tolerant quantum systems can provide exponential-time speedup in problems that are considered to be classically hard, such as quantum simulation of dynamical systems ,\cite{Cao_2019,Olsen_etal_2017} 
and quantum information processing tasks such as factorization  \cite{Zoller_2005,Shor_1997}
technological limitations originating primarily from the inherently unstable nature of quantum systems restrict the capability of current quantum processors. Additionally, the large number of physical qubits needed to ensure the correctness of even a single logical qubit to a high degree of certainty given currently available quantum error correction codes requires large-scale quantum processors to perform accurate computations. These unsolved challenges have led the current era to be termed the noisy intermediate scale quantum (NISQ) era by Preskill \cite{Preskill_2018}. 

\begin{figure}[h]
\centerline{\includegraphics[width=0.5\textwidth]{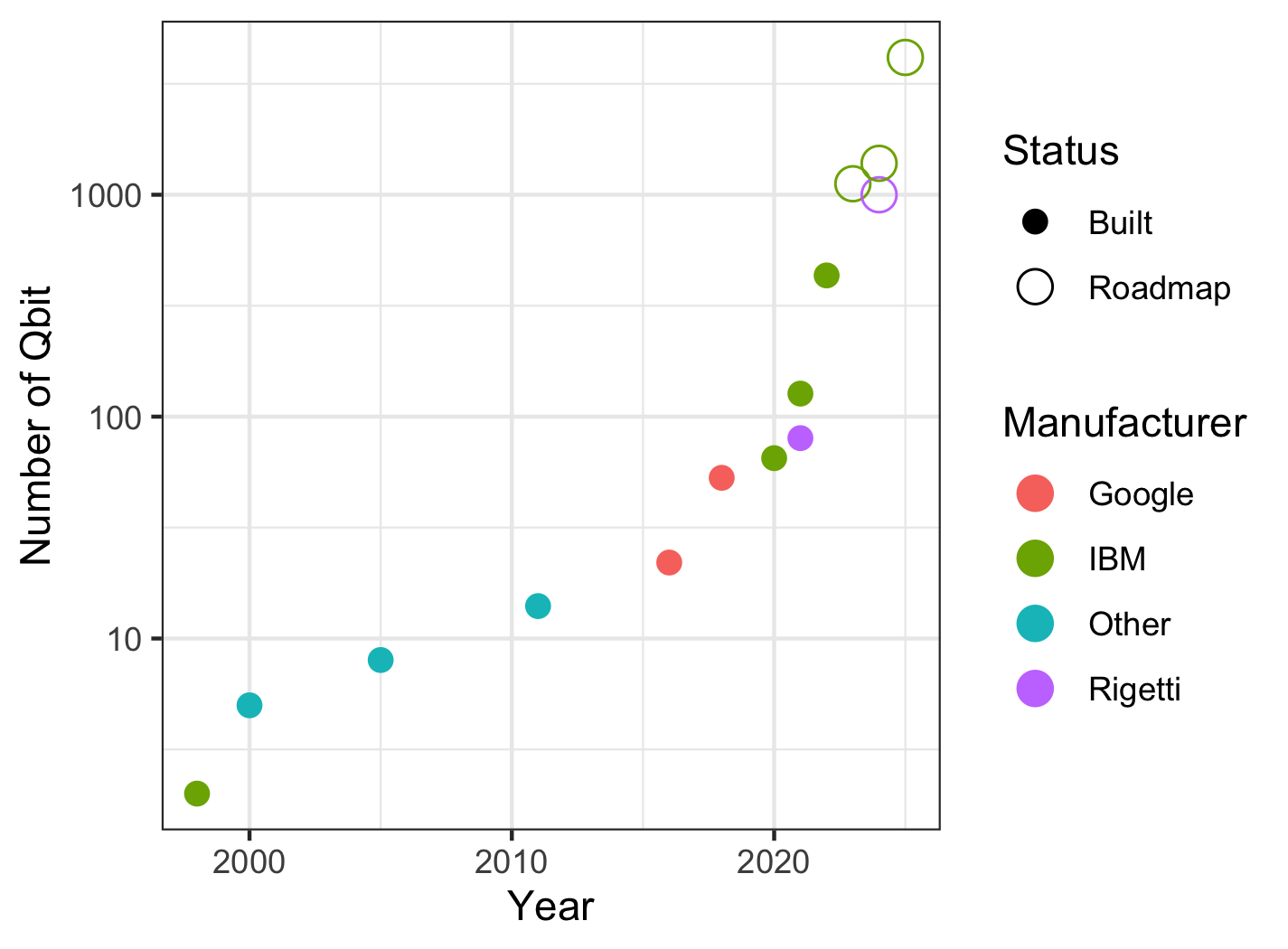}}
\caption{Number of qubit for quantum computers based on the year of development. Published hardware roadmaps indicate that significantly larger quantum computers are likely to  become available within the next few years.}
\label{fig:qubit-growth}
\end{figure}

Current NISQ processors, while serving as fascinating testbeds for small-scale experimentation, are unlikely to produce practical results in the short term. However, the growth rate in recent years of the number of qubits contained in quantum processors \ref{fig:qubit-growth} is reason to believe we will continue to move towards processors able to provide quantum advantage on a wide variety of tasks.


To be able to effectively use techniques such as Shor's polynomial-time factoring algorithm 
\cite{Shor_1997}
in applications such as RSA cryptography, which is estimated to require on the order of 4,098 logical qubits \cite{Sevilla_2020,Haner_2016}
progress needs to be made in multiple areas, namely algorithmic efficiency, error-correcting code efficiency, and physical qubit noise. Algorithmic improvements on the original algorithm, such as 
\cite{Cherckesova_2020,Leander_2002}
may lead to fewer necessary logical qubits to achieve the same factorization speedup. Additionally, the discovery of better error-correction codes as well as improvements made to physical qubits on the hardware end would lead to fewer physical qubits being required to some similar classes of problems. 


Some estimates suggest that O(100,000) qubit quantum computers will be required to serve as practical tools to threaten cryptosystems \cite{Popescu_2022}. Based on the trajectories depicted in Figure \ref{fig:qubit-growth}, it conceivable that quantum computers this size will be built within the next decade, and possibly even sooner. Since complexity-based cryptography has been the dominant security paradigm for several decades, this trajectory presents problems for other classes of cryptographic algorithm as well. Table \ref{tab:post-quantum} summarizes conclusions of a 2016 report from the National Institute of Standards and Technology (NIST) regarding how quantum computers can be expected to impact major cryptographic algorithms \cite{Chen_2016}.
These projections assume that more mature quantum computers will be developed, 
which are generally constrained by DiVincenzo's criteria with respect to physical realization of the system \cite{DiVincenzo}. We may consider two criteria in particular as having particular importance for the NISQ era: (1) establishing scalability for the physical system; and (2) long decoherence times relative to the operation time for quantum gates. Developing quantum computers that can demonstrate these capabilities is a key engineering challenge for the next decade. While the trajectory shown in Figure \ref{fig:qubit-growth} may seem ominous from the perspective of information security, simply adding qubits is not sufficient to satisfy the requirement for scalability. Since, measurement capabilities and device noise properties can change with system size, the only way to accurately project performance capabilities is to run algorithms. Due to noisy nature of current hardware, quantum algorithms can be expected to significantly underperform expectations in the context of Figure \ref{fig:qubit-growth}.
Developing more realistic assessments requires us to understand how individual algorithms actually perform under the influence of noise, and potentially even use this information to develop algorithms that are relatively more noise tolerant. 

\begin{table}[htbp]
\label{tab:post-quantum}
\caption{Impact of quantum computing on major cryptographic algorithms.}
\begin{center}
\begin{tabular}{|P{2.2cm}|P{2cm}|P{2.5cm}|}
\hline
\textbf{Cryptographic Algorithm} &  \textbf{Purpose} &
\textbf{Post-Quantum Requirement}\\
\hline
AES & Encryption & Larger key sizes \\
\hline
SHA-2,SHA-3 & Hash functions & Larger output \\
\hline
RSA & Signatures, key establishment & Not secure \\
\hline
ECDSA, ECDH & Signatures, key exchanges & Not secure\\
\hline
DSA & Encryption & Not secure\\
\hline
\multicolumn{3}{l}{Adapted from Chen et al. \cite{Chen_2016}}
\end{tabular}
\end{center}
\end{table}

Table \ref{tab:shor} demonstrates the hypothesized timing for a classical computer and a quantum computer to break an RSA encryption. Due to existing quantum algorithms and parallelism, the integrity of a 256-bit key on a classical computer has a computational equivalent of a 128-bit key against a quantum attack. It was found that a classical computer would need about 6 quadrillion years to break a 2048-bit RSA encryption while quantum computers have the potential to break the key within a few hours \cite{Lindsey_2020}. Similarly, a 4,096-bit key is deemed unbreakable in a classical computer due to needing multiple lifetimes to find the solution, but quantum computers can break it within a matter of hours using Shor’s algorithm \cite{Denning_2019}. Given a $N$-bit RSA key, Shor's algorithm can factor the key in time 
O($(\log N)^2(\log \log N) (\log \log \log N) $) \cite{Shor1994}. Assuming a baseline time of $\sim 3$ hours to factor a 2048-bit RSA key on a hypothetical future quantum computer, Table \ref{tab:shor} shows how this scaling would be expected to influence factorization times for larger keys. From these results it is clear that larger RSA keys do not provide a reasonable degree of protection in a post-quantum world. The potential for computational capabilities of quantum computing that surpass the abilities a traditional computer motivates a clear need for improved quantum-resistant cryptographic approaches.

\section{Methods}
\label{sec:methods}

\subsection{Co-Design and the cuQuantum Platform}

In comparison with classical computers, current quantum computers are rare, unreliable and expensive. 
In spite of the rapid developments that have been occurring with respect to quantum computing,
quantum systems do not currently provide computational advantages relative to classical hardware. 
Within this context, theoretical projections represent the upper bound for what quantum computers may
eventually be able to do. Realistically assessing the actual capabilities requires direct experimentation.
Software co-design can provide an important mechanism to assess hardware maturity as well as understand how
hardware design factors influence the computational bottlenecks that arise in practice. The idea to ``co-design'' 
computing systems is driven by this feedback loop; information about practical computational bottlenecks is 
particularly valuable as a way to improve hardware design. As quantum computers become increasingly capable,
this more practical approach will become essential to understand where cryptographic algorithms actually 
stand. Informed strategies to estimate risks to cryptography will need to rely on practical measurements
as well as theoretical estimates. 

\begin{figure}[h]
\centerline{\includegraphics[width=0.5\textwidth]{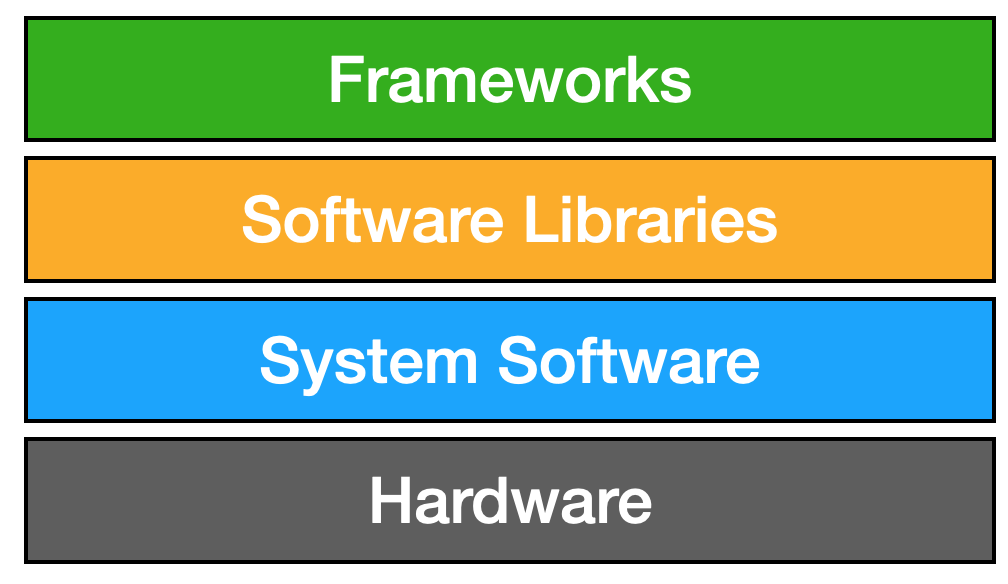}}
\caption{Summary of software components and their relationship to hardware
in the context of co-design. Most users write code in programming frameworks (cirq, qiskit, etc.)
that at sit at the top layer. Software libraries and system software are essential to translate
these codes so that they can run efficiently on hardware. }
\label{fig:software}
\end{figure}

The capacity for quantum computers to become useful tools is heavily constrained
by the maturity of associated software environments. Conceptually, software dependencies for
quantum computing can be considered in the simplified context of Figure \ref{fig:software}. 
Typical users write code in programming frameworks such as {\em cirq} and {\em qiskit}, which 
provide high-level abstractions for quantum gates, and are typically developed by hardware
vendors as a mechanism to reduce the barrier to entry for programming quantum devices. 
Ultimately, lower level software tools are needed to convert high-level programs into a form that
be associated with physical measurements taken from the hardware. While the associated software ecosystems can become relatively complex, for our purposes it is sufficient to consider two abstractions 
that carry out this mapping: software libraries and system software. The purpose of software libraries is to provide common abstractions that can encapsulate common requirements of general quantum algorithms as they are typically expressed at the framework layer. That is, they need to capture the information in a quantum circuits and re-express the algorithm in a way that can map more efficiently to the underlying hardware. The system software layer provides another layer of abstraction, and is responsible for applying measurements performed on the physical system to the inputs defined from the software. Well-designed software systems are able to insulate the programmer from hardware details so that the programs are minimally impacted by changes at the hardware layer. The process of software co-design is essentially constructed to streamline the design of these components so that the overall system can function as well as possible. 

In this work, our strategy relies on using the {\em cuQuantum} package developed by NVIDIA. 
In the context of Figure \ref{fig:software}, {\em cuQuantum} is a software library that 
includes efficient data structures and routines to perform vector and tensor operations that 
arise in many quantum programs. The basic structures supported by the API are {\em cuStateVec} 
and {\em cuTensorNet}, which can be referenced from either {\em C} or {\em python}. Common frameworks
supported by {\em cuQuantum} include both {\em cirq} (google) and {\em Qiskit} (IBM), which are python-based interfaces. In the current era of quantum computing, the utility of {\em cuQuantum} is associated with the ability to support both classical computing (e.g. CPU and GPU) and quantum computing so that the same programs can be run efficiently on both classes of computing system. Of existing classical computers, GPU are among the more powerful general-purpose computing instruments. Quantum programs targeting GPU-based systems are currently able to simulate quantum programs that too large run on 
current NISQ devices. In particular, programs that require large circuit depth (such as Shor's algorithm) include important situations where classical computing can be applied as a tool to advance algorithms and software to improve computational readiness and software maturity for future quantum computers. Furthermore, classical computers are far more reliable and less noise-prone 
and can provide a way to conduct controlled studies on how circuit noise influences different algorithms.
We will apply {\em cuQuantum} to simulate several important quantum circuits on widely-available 
commodity GPU hardware, and also develop basic methods to study how noise influences the reliability of these algorithms. 

\subsection{GHZ circuit}
\label{sec:GHZ}

\begin{figure}[h]
\centerline{\includegraphics[width=0.5\textwidth]{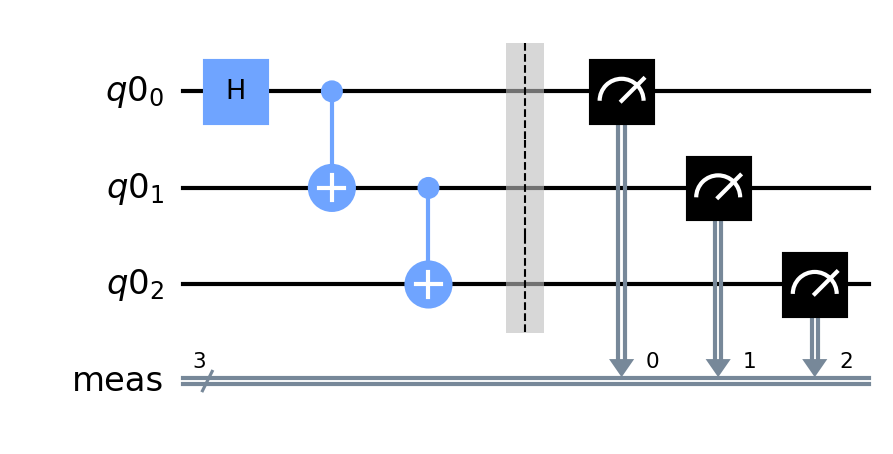}}
\caption{The GHZ circuit is constructed by combining Hadamard and Controlled Not gates to create a state of maximum entanglement. Arrow symbols denote when quantum measurements are performed. }
\label{fig:GHZ}
\end{figure}

The GHZ circuit was first devised as an approach to experimentally demonstrate entanglement \cite{GHZ_1990}. Since the physics of entanglement are central to the construction of a general-purpose quantum computer, the GHZ circuit is commonly used to assess their ability to measure desired physical behaviors. Figure \ref{fig:GHZ} shows the GHZ circuit, which illustrates several key elements of quantum circuits. qubits are listed along the vertical direction; in this case, there are three qubits, with the bottom line used to denote when physical measurements are performed based on the circuit. Each qubit associated with a state vector $|\psi_q\big>$ that is a superposition of two possible states.
The state of a system with $Q$ qubits will therefore be described by a state vector with $2^Q$ elements.
Information flows from left to right, passing through a series of quantum gates that operate on the qubit in various ways. Quantum gates are represented by unitary matrices, meaning that the norm of the input 
vector will be preserved when the operation is applied. The gates define the expected behaviors based on the circuit representation. On a physical quantum machine, the behavior of the gates may deviate 
from what is expected due to the noise associated with the quantum measurement process. The addition of noise to emulated circuits will be described in \S \ref{sec:noise}. 

The basic set of gates used to construct the GHZ circuit are the Hadamard gate and the  ``Controlled Not'' gate. 
The Hadamard gate operates on a single qubit, and is therefore represented by a $2 \times 2$ matrix,
\begin{equation}
\mbox{H} \equiv \frac{1}{\sqrt{2}}\begin{pmatrix} 1 & 1 \\ 1 & -1\end{pmatrix}
\label{eq:Hadamard}
\end{equation}
The controlled not gate operates on two qubits, and is therefore represented by a $4\times 4$ matrix,
\begin{equation}
\mbox{CX} \equiv 
\begin{pmatrix}
1 & 0 & 0 & 0 \\ 
0 & 1 & 0 & 0 \\ 
0 & 0 & 0 & 1 \\ 
0 & 0 & 1 & 0 
\end{pmatrix}
\label{eq:CNOT}
\end{equation}

The GHZ circuit is constructed by applying the Hadamard gate to each qubit, followed by a series of CNOT gates to entangle the qubits. This results in the maximally entangled state known as the GHZ state, which is in a superposition of all qubits being 0 or all qubits being 1.

\begin{equation}
|GHZ\big> = \frac{|000\big> + |111\big>}{\sqrt{2}}
\label{eq:GHZ}
\end{equation}
When a measurement is applied to the qubits, the result is $\ket{000}$ or $\ket{111}$ with equal probability. The purpose of the circuit is to demonstrate the capability for a hardware device to exhibit the phenomenon of quantum entanglement.

\subsection{Hidden Shift Algorithm}
\label{sec:HS}

Hidden shift problems represent a broad class of quantum algorithms that have important consequences for cryptography \cite{Dam_2002}. The problem may be stated as follows: suppose we have two functions $f$ and $g$ such that the property $f(x) = g(x+s)$ holds for every $x$ based on some unknown shift $s$. The objective of the algorithm is to find $s$. This problem is relevant to RSA encryption because it encapsulates important algorithms used to generate pseudo-random numbers. On a classical computer, the task of finding $s$ is computationally intractable. On a quantum computer, the task can be solved in polynomial time which poses a risk to certain cryptosystems.

\begin{figure}[h]
\centerline{\includegraphics[width=0.5\textwidth]{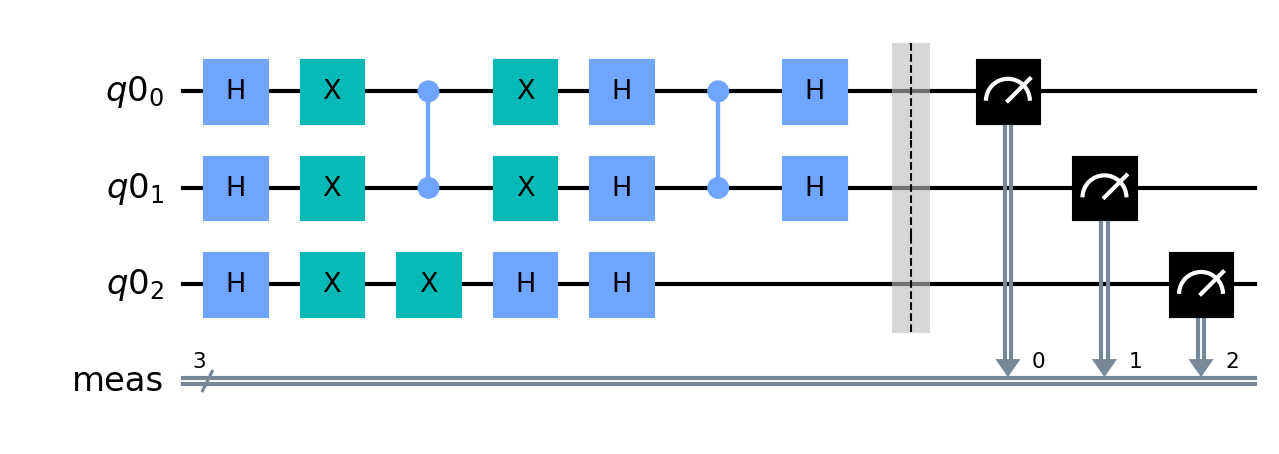}}
\caption{Hidden Shift Quantum Circuit}
\label{fig:HS}
\end{figure}

\noindent \textbf{Hidden Shift Quantum Algorithm}
\begin{enumerate}
\item Prepare the quantum state in the initial state $\ket{0}^N$
\item Make a superposition of all inputs $\ket{\psi_x}$ with a set of Hadamard gates
\item Compute the shifted function $f(x) = g(x \oplus s)$ into the phase with a proper set of gates
\item Apply a Fourier Transform to generate another superposition of states with an extra phase that is added to $g(x \oplus s)$
\item Query the oracle $g$ into the phase with a proper set of controlled gates
\item Apply another set of Hadamard gates which act now as an Inverse Fourier Transform to get the state $\ket{s}$
\item Measure the resulting state to get $s$
\end{enumerate}

\subsection{Noise Emulation in Quantum Circuits}
\label{sec:noise}

A useful advantage of running simulations of quantum circuits on commodity hardware is the ability to explicitly control the influence of noise on the output. On real NISQ hardware, noise is an unavoidable consequence of the system design. Today, we do not have the ability to build quantum devices that achieve the low-levels of noise and error correction needed to support practical computations. From the hardware perspective, improving the noise properties of quantum devices is an important challenge for research and development. From the software perspective, it is useful to understand how noise influences algorithm behavior, since this is likely to determine how soon an algorithm can benefit from a real quantum computer. We can easily emulate quantum computers that have significantly better noise characteristics in comparison to physical NISQ devices. This motivates our decision to rely on emulation and develop methods to perform such characterizations.

Intuitively, noise is introduced into a circuit by mixing information from the actual state vector with random information. A common model to accomplish this is with a quantum depolarizing channel,  which is used to model the effects of noise on a quantum system \cite{King_depolarize}. It is constructed by randomly applying one of four unitary operations to a qubit, each with a certain probability. We use {\em cirq} to introduce depolarizing channels into quantum circuits based on the {\tt cirq.depolarize} function, which transforms the density matrix $\rho$ based on the Pauli gates $P_i$
\begin{equation}
    \rho \leftarrow  (1-p) \rho + \frac{p}{4^n - 1} \sum_{i}^{n-1} P_i \rho P_i \;,
\end{equation}
where $n$ is the number of qubits. The first term corresponds to the original state of the quantum
system. The second term corresponds to the noise added for each of the gates. The probability $p$ determines the amount of noise that is added. It is clear that
$p=0$ will correspond to the situation where the circuit is noise-free, and $p=1$ will correspond to the situation where the output will be determined entirely from noise. 
The Pauli gates are typically used in depolarizing channels as they are the most basic unitary operations that can be applied to a qubit. These operations are the identity, the Pauli X, Y, and Z gates, represented by the 2x2 matrices:

\[I = \begin{pmatrix} 1 & 0 \\ 0 & 1
\end{pmatrix}\]
\[ \sigma_X \equiv \begin{pmatrix} 0 & 1 \\ 1 & 0 \end{pmatrix}\]
\[ \sigma_Y \equiv \begin{pmatrix} 0 & -i \\ i & 0 \end{pmatrix}\]
\[ \sigma_Z \equiv \begin{pmatrix} 1 & 0 \\ 0 & -1 \end{pmatrix}\]

The identity operation is applied with probability $1-p$, while each of the Pauli gates is applied with probability $p/3$. This has the effect of randomly flipping the qubit's state with probability $p$, and leaving it unchanged with probability $1-p$.

In our algorithm, we added a depolarizing channel after every time slice of quantum gates in the circuit in order to emulate noise in the simulations. This enabled us to model the effects of quantum states deteriorating over time, an important issue in the current NISQ landscape. 



\section{Results}


In this section we evaluate the performance for quantum algorithms described in \S \ref{sec:methods}  based on state-of-the-art classical computing resources. The {\em cuQuantum} environment 
was deployed on NVIDIA DGX 100 servers based on the {\em TinkerCliffs} computing system at 
Virginia Tech. Each DGX node is equipped with $8 \times$ NVIDIA A100 GPU and 128 CPU
processor cores (AMD EPYC 7742). Each A100 GPU has 80 GB of memory, and the total CPU memory within a node is 2048 GB. Memory determines the maximum problem size that can be simulated, which effectively constrains the number of qubits that can be simulated by the hardware. Larger supercomputing systems also exist, which can support hundreds or even thousands of GPU servers linked together with high performance interconnects. This type of infrastructure can support larger problem sizes by distributing the memory required to represent qubits across the memory for many different physical computers. Since configurations with up to eight GPU are relatively common in commercial cloud environments, our testing setup is consistent with computational resources that are broadly accessible to the cybersecurity research community.

\begin{figure}[h]
\centerline{\includegraphics[width=0.5\textwidth]{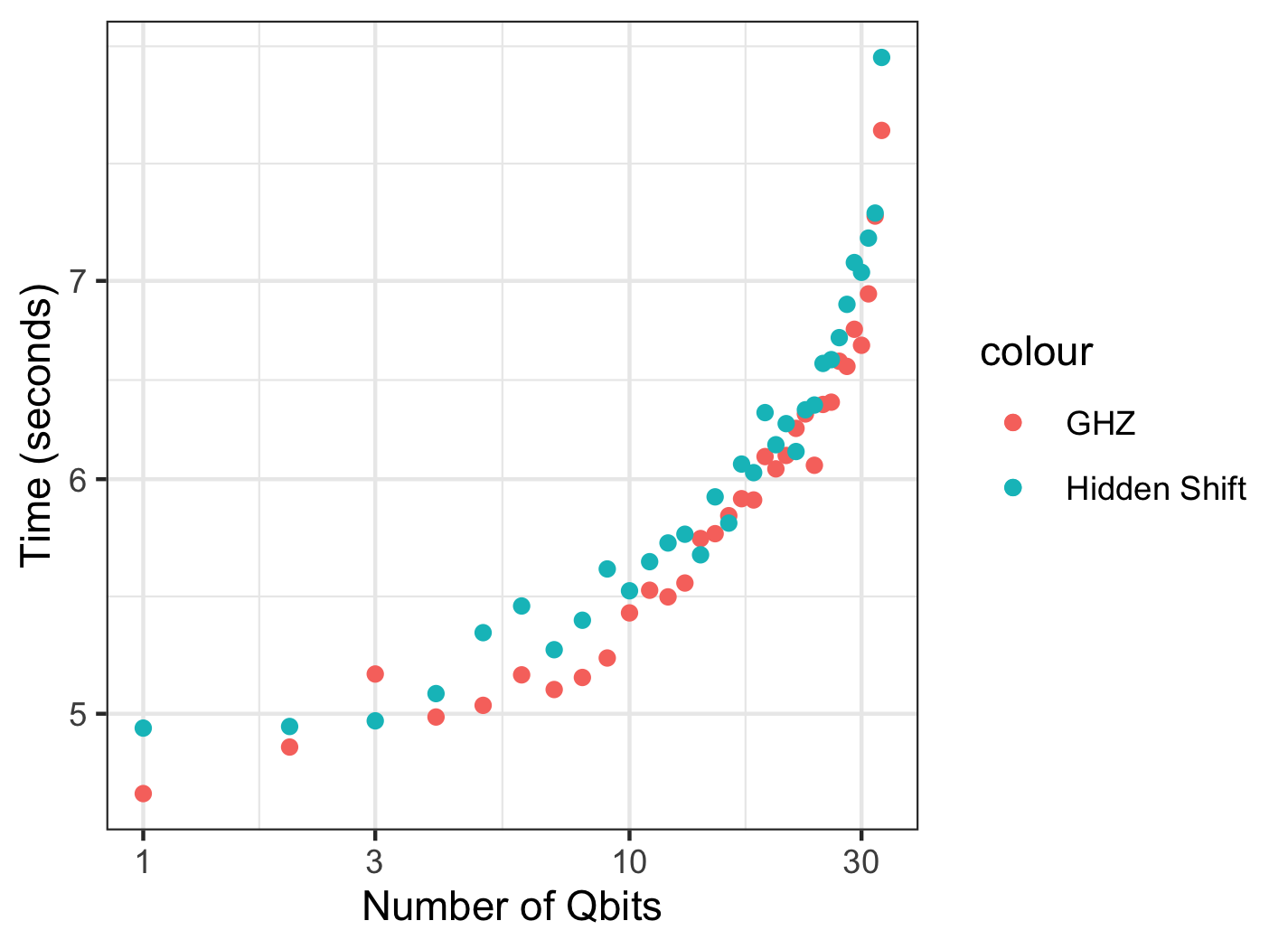}}
\caption{Computational time required to execute {\em GHZ} and {\em hidden shift} algorithms on NVIDIA A100 GPU using {\em cuQuantum}. }
\label{fig:qubit-timing}
\end{figure}

The computational time required to simulate quantum algorithms depends on the size of the quantum circuit and the number of qubits. Generally, the larger the quantum circuit and the more qubits, the longer the computational time. Although quantum algorithms are developed, since simulations are performed on a classical computer, the performance and scaling behavior are determined accordingly.
Simulating a quantum circuit with $n$ qubits requires $2^n$ vectors. For example, a quantum circuit with 10 qubits requires vectors of length 1024. As seen in Figure \ref{fig:qubit-timing} the time required to simulate a quantum circuit increases exponentially with the number of qubits. The bottleneck in the simulations resulted being the amount of memory required to store the quantum state vector. As the number of qubits increases, the amount of memory required to store the quantum state vector increases exponentially. We were able to simulate circuits with up to 32 qubits on a single A100 GPU. Due to the limited circuit depth, the time required to simulate the GHZ and hidden shift algorithms is typically less than 10 seconds. Similar computational costs are associated with the two algorithms. We note that due to the implementation of the {\em cirq} noise model, lower performance was obtained when the noise model was imposed on the circuit. 


To visualize the effects of noise on quantum circuits, we implemented a noisy version of the hidden shift algorithm using {\tt cirq.depolarize} as described above. Holding the number of samples constant, we ran our noisy implementation of hidden shift at different noise levels thirty times, plotting the number of times we successfully discovered the correct shift amount. As increased circuit complexity leads to more noisy operations being performed and hence a greater likelihood of unrecoverable state collapse, we also hypothesized that algorithmic accuracy decreases with a higher number of qubits. To test this theory, we repeated our process with different numbers of qubits.


\begin{figure}[h]
\centerline{\includegraphics[width=0.5\textwidth]{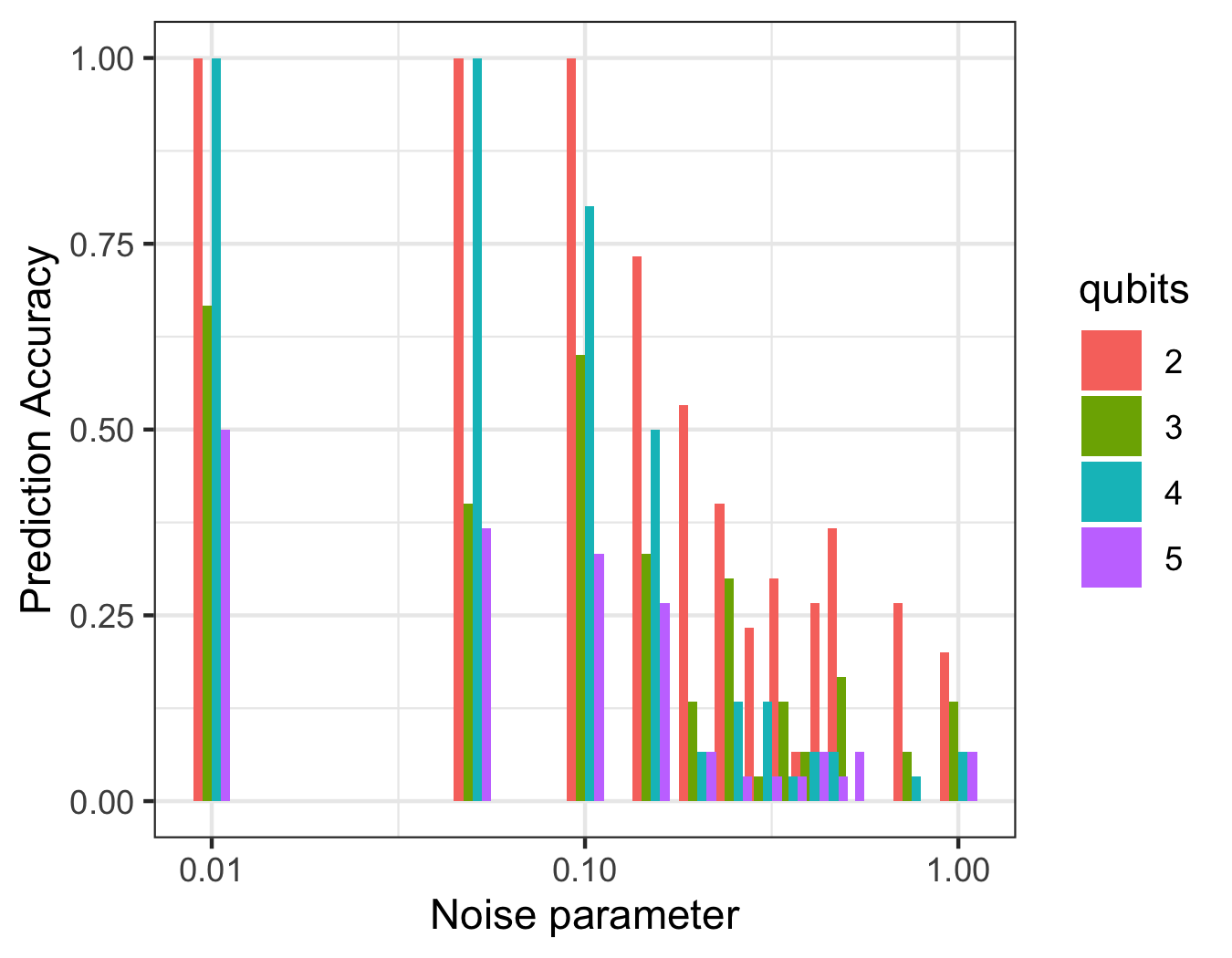}}
\caption{Effects of noise on the accuracy of the hidden shift algorithm with varying number of qubits. }
\label{fig:noisy-hs}
\end{figure}

Our results are displayed in Figure \ref{fig:noisy-hs}, which shows the prediction accuracy vs noise for various number of qubits. Increased circuit complexity has an apparent effect on the robustness of quantum circuits to noise, with the algorithms with fewer qubits performing significantly better than the algorithms with more qubits. This is expected, since the probability to guess the correct bit sequence diminishes with the number of qubits. In addition, as expected, noisier circuits on average lead to worse performance. 

\section{Summary \& Conclusions}


In this paper we evaluate quantum algorithms on classical hardware using the {\em cuQuantum} framework. The advantage of this approach is that it can be used to design and test codes that can run efficiently on both classical and quantum hardware. We demonstrate that this approach can be used to simulate quantum circuits with up to 32 qubits using a single NVIDIA A100 GPU, which is sufficient to perform meaningful experiments with respect to the impact of noise on quantum algorithms. For circuits without noise, simulations make efficient use of GPU resources and are able to fully utilize the {\em cuQuantum} library to perform emulation. A quantum noise channel was modeled using the {\em cirq} framework, which was used to assess the breakdown of the hidden shift algorithm in situations where gate noise influences the accuracy of the results.  We note that reduced computational efficiency was observed when the noise model was enabled,
due to the fact that this particular tool does not use the GPU as efficiently as other 
gates in the circuit. However, given that the structure of the noise model is modeled based on 
the Pauli gates, it should be possible to develop more efficient approaches for noise emulation 
by directly imposing these components within the circuit. Furthermore, it will be important to 
generalize the approach to consider other noise models to account for different underlying physical causes of gate noise within hardware designs. 

A clear understanding of how noise influences cryptographic algorithms can inform projections for future risk in the context of cybersecurity. Theoretical predictions represent a worst-case scenario, considering the case of a perfectly efficient quantum computer that is not hindered by hardware bottlenecks. Adopting a software co-design approach to cryptographic algorithms can provide additional practical information, seeking to identify the most likely scenarios. Codes developed in this fashion can also be used to formulate unbiased metrics for hardware devices. Quantum emulation can be used to assess a wide range of noise scenarios, ranging from very low to very high noise.
The associated benchmarks can provide information on hardware maturity that captures appropriate context for particular algorithms of concern. The preliminary efforts described in this work suggest that emulation is likely to play an important role in the evaluation of early-stage NISQ devices, as well as the development of associated software capabilities. 

\section*{Acknowledgment}
This work was supported by the Senior Military College Cyber Institute (SMC21) and the Hume Center for National Security and Technology. The authors acknowledge Advanced Research Computing at Virginia Tech for providing computational resources and technical support that have contributed to the results reported within this paper.

\end{document}